\documentclass[%
 aip,
 amsmath,amssymb,
 reprint,%
]{revtex4-1}
\usepackage{graphicx}% Include figure files
\usepackage{dcolumn}% Align table columns on decimal point
\usepackage{bm}% bold math
\usepackage{color}
\usepackage[utf8]{inputenc}
\usepackage[T1]{fontenc}
\usepackage{mathptmx}
\usepackage{multirow}
\usepackage{float}
\usepackage{epsfig}
\usepackage{epstopdf}

\usepackage{url}
\usepackage{titlesec}
\titleclass{\subsubsubsection}{straight}[\subsection]
\newcounter{subsubsubsection}[subsubsection]
\renewcommand\thesubsubsubsection{\thesubsubsection.\alph{subsubsubsection}}
\titleformat{\subsubsubsection}
 {\normalfont\normalsize\bfseries}{\thesubsubsubsection}{1em}{}
\titlespacing*{\subsubsubsection}
{0pt}{3.25ex plus 1ex minus .2ex}{1.5ex plus .2ex}
\begin{document}

\preprint{AIP/123-QED}

\title{Effects of the initial perturbations on the Rayleigh-Taylor-Kelvin-Helmholtz instability system}
% Force line breaks with \\

\author{Feng Chen } %\begin{CJK*}{GBK}{song}(�·�)\end{CJK*}}
 \thanks{Corresponding author: chenfeng-hk@sdjtu.edu.cn, shanshiwycf@163.com}
 \affiliation{School of Aeronautics, Shan Dong Jiaotong University, Jinan 250357, China.}

\author{Aiguo Xu}% \begin{CJK*}{GBK}{song}(������)\end{CJK*}}%
 %\email{Corresponding author. Email addresses: Xu\_Aiguo@iapcm.ac.cn}
 \thanks{Corresponding author: Xu\_Aiguo@iapcm.ac.cn}
\affiliation{Laboratory of Computational Physics, Institute of Applied Physics and Computational Mathematics, P. O. Box 8009-26, Beijing 100088, China.}
\affiliation{HEDPS,Center for Applied Physics and Technology, and College of Engineering, Peking University, Beijing 100871, China.}
\affiliation{State Key Laboratory of Explosion Science and Technology, Beijing Institute of Technology, Beijing 100081, China.}

\author{Yudong Zhang} %\begin{CJK*}{GBK}{song}(������)\end{CJK*}}
%\homepage{http://www.Second.institution.edu/~Charlie.Author.}
\affiliation{School of Mechanics and Safety Engineering, Zhengzhou University, Zhengzhou 450001, China.}

\author{Yanbiao Gan} %\begin{CJK*}{GBK}{song}(���ӱ�)\end{CJK*}}
\affiliation{Hebei Key Laboratory of Trans-Media Aerial Underwater Vehicle, North China Institute of Aerospace Engineering, Langfang 065000, China.}

\author{Bingbing Liu} %\begin{CJK*}{GBK}{song}(������)\end{CJK*}}
\affiliation{Naval Architecture and Port Engineering College, Shan Dong Jiaotong University, Weihai 264200, China.}

\author{Shuang Wang} %\begin{CJK*}{GBK}{song}(��ˬ)\end{CJK*}}
\affiliation{School of science, Shandong Jianzhu University, Jinan 250101, China.}
\date{\today}% It is always \today, today,
             %  but any date may be explicitly specified
\begin{abstract}
In the paper, the effects of initial perturbations on the Rayleigh-Taylor instability (RTI), Kelvin-Helmholtz instability (KHI), and the coupled Rayleigh-Taylor-Kelvin-Helmholtz instability (RTKHI) systems are investigated using a multiple-relaxation-time discrete Boltzmann model. Six different perturbation interfaces are designed to study the effects of the initial perturbations on the instability systems. It is found that the initial perturbation has a significant influence on the evolution of RTI. The sharper the interface, the faster the growth of bubble or spike. While the influence of initial interface shape on KHI evolution can be ignored.
Based on the mean heat flux strength $D_{3,1}$, the effects of initial interfaces on the coupled RTKHI are examined in detail. The research is focused on two aspects: (i) the main mechanism in the early stage of the RTKHI, (ii) the transition point from KHI-like to RTI-like for the case where the KHI dominates at earlier time and the RTI dominates at later time. It is found that the early main mechanism is related to the shape of the initial interface, which is represented by both the bilateral contact angle $\theta_{1}$ and the middle contact angle $\theta_{2}$. The increase of $\theta_{1}$ and the decrease of $\theta_{2}$ have opposite effects on the critical velocity. When $\theta_{2}$ remains roughly unchanged at $90$ degrees, if $\theta_{1}$ is greater than $90$ degrees (such as the parabolic interface), the critical shear velocity increases with the increase of $\theta_{1}$, and the ellipse perturbation is its limiting case; If $\theta_{1}$ is less than $90$ degrees (such as the inverted parabolic and the inverted ellipse disturbances), the critical shear velocities are basically the same, which is less than that of the sinusoidal and sawtooth disturbances. The influence of inverted parabolic and inverted ellipse perturbations on the transition point of the RTKHI system is greater than that of other interfaces: (i) For the same amplitude, the smaller the contact angle $\theta_{1}$, the later the transition point appears; (ii) For the same interface morphology, the disturbance amplitude increases, resulting in a shorter duration of the linear growth stage, so the transition point is greatly advanced.
\end{abstract}

\maketitle

\section{Introduction}

Hydrodynamic instability-driven mixing processes are ubiquitous in nature, and of great significance both for fundamental research and for engineering application in a number of research fields, including Inertial Confinement Fusion (ICF), astrophysical phenomena, and supersonic combustion. In the last several decades, much attention has been paid to the challenging problem
\cite{Zhou2017PR-I,Zhou2017PR-II,Zhou2019PP,LiHF2021PRE,WangLF2009POP,epl2010,Liang2016PRE,Liang2019POF,Liang2021POF,Zou2017JMES,Zou2017PRE,Zou2019JFM,RavidNatureC,chen2021jfm,Lai2020,Lai2021,Tang2021,LL2019,Tao2020,Tao2021}.

In view of its direct consequence in various applications, the effects of initial perturbations are being explored extensively
\cite{Miles2004,Dimonte2004,Ramaprabhu2005,Olson2009,Gowardhan2011,Doron2011,Wei2012,Kuchibhatla2013,Liu2013,McFarland2013,ZhaiJFM2014,ZhaiJFM2015,Dell2015,Xiao2016,Xie2017,Kord2019,Ding2020,Si2021}.
Miles et al. \cite{Miles2004} investigated the effect of initial conditions on two-dimensional (2D) Rayleigh-Taylor instability (RTI) and transition to turbulence in planar blast-wave-driven systems, and found that the initial conditions have a strong effect on the time to transition to the quasi-self-similar regime. Dimonte \cite{Dimonte2004} studied the dependency of the self-similar Rayleigh-Taylor bubble acceleration constant on the initial perturbation amplitude. Ramaprabhu et al. \cite{Ramaprabhu2005} investigated the dependence of the RTI growth coefficient and the self-similar parameter on the amplitude, the spectral shape, the longest wavelength imposed, and mode-coupling effects. Gowardhan et al. \cite{Gowardhan2011} demonstrated that the initial material interface morphology controls the evolution characteristics of Richtmyer-Meshkov instability (RMI). Wei et al. \cite{Wei2012} investigated the role of the initial perturbation shape on the RTI development. It is found that the time when the instability reaches a mean quadratic growth depends on the initial perturbation shape, and the subsequent vortical interactions are also very sensitive to details of the initial perturbation shape. Liu et al. \cite{Liu2013} found that the temporal evolution of the bubble tip velocity is sensitively dependent on the Atwood numbers, the initial perturbation amplitude and the initial perturbation velocity in classical RTI. McFarland et al. \cite{McFarland2013} investigated the effects of inclination angle and incident shock Mach number on the inclined interface RMI. Zhai
et al. studied the RMI problems of three light gas interfaces \cite{ZhaiJFM2014} and six heavy gas interfaces \cite{ZhaiJFM2015}, including the different shock wave refraction on the interface and the influence of initial interface shape on the interface features. Dell et al. \cite{Dell2015} systematically studied the effect of the initial perturbation on RMI, and observed that the initial growth rate of RMI is a non-monotone function of the initial perturbation amplitude. Xiao et al. \cite{Xiao2016} studied the effects of an initial perturbation on RMI, and found that the evolution of the interface with large initial amplitude in a low-density nonuniform area is fastest, while that with a small initial amplitude in a high-density nonuniform area is slower. Xie et al. \cite{Xie2017} showed that the growth of the RTI mixing zone is substantially retarded by superimposing an optimized additional mode on its random initial perturbations. Liang et al. \cite{Liang2019POF} analyzed the effects of the initial conditions in terms of the perturbation wavelength and amplitude. It is found that the instability undergoes a faster growth at the intermediate stage for a larger wavelength, while the late-time bubble and spike growth rates are insensitive to the changes of the initially perturbed wavelength and amplitude. Kord et al. \cite{Kord2019} investigated the sensitivity of objective functions to the initial perturbation amplitudes at various stages of the RTI, such as the mole fraction, kinetic energy norm. Ding et al. \cite{Ding2020} demonstrated that controlling the layer thickness is an effective way to modulate the late-stage instability growth, which may be useful for the target design. Si et al. \cite{Si2021} investigated the mode-composition effect on the three-dimensional (3D) interface RMI development, and found that the mode-coupling has an evident influence on the bubble evolution.

It is generally believed that in the ICF, in the late stage of RTI, the materials on both sides penetrate each other to form shear flow, and Kelvin-Helmholtz instability (KHI) occurs at the interface. However, in nature or practical engineering applications, such as in the atmosphere and oceans, air/fuel mixing in combustion chambers, the outer region of supernovae, and the compression of the fuel capsule in ICF, there is rarely a single instability. Due to the complexity of the interface, KHI often occurs simultaneously when RTI develops. It would be interesting to study the effect of the combined occurrence of two instability mechanisms. A complete understanding of the relation between the RTI and the KHI is important in understanding the hydrodynamic instability-driven mixing processes. In addition to the growth rate and final state, some studies have been made on the early linear and early nonlinear evolution of the coupled instability \cite{Wang2010,Ye2011,Mandal2011,Olson2011,Akula2013,Vadivukkarasan2016,Vadivukkarasan2017,Vadivukkarasan2020,Sarychev2019,Brizzolara2021,Chen2020}.
Both the instability modes and devolution processes that may be manifested when two such mechanisms occur simultaneously are of current interest. Ye et al. \cite{Wang2010,Ye2011} investigated the competitions between RTI and KHI in 2D incompressible fluids within a linear growth regime. It is found that the competition between the RTI and the KHI is dependent on the Froude number, the density ratio of the two fluids, and the thicknesses of the density transition layer and the velocity shear layer. Mandal et al. \cite{Mandal2011} investigated the nonlinear evolution of bubble and spike due to the combined action of RTI and KHI.
Olson et al. \cite{Olson2011} studied the coupled RTKHI in the early nonlinear regime, and indicated a complex and non-monotonic behavior where small amounts of shear in fact decrease the growth rate.
Akula et al. \cite{Akula2013} showed that the superposition of shear on RTI at small Atwood number increases the mixing width and growth rate at early times. Vadivukkarasan et al. \cite{Vadivukkarasan2016,Vadivukkarasan2017,Vadivukkarasan2020} described the 3D destabilization characteristics of cylindrical and annular interfaces under the combined RTKHI,
and the effects of various parameters on the most unstable wavenumbers were studied.
Sarychev et al. \cite{Sarychev2019} found that an undulating topography on the interface coating/base material is resulted from the coupled RTKHI. Brizzolara et al. \cite{Brizzolara2021} observed that the cross-over time can correctly predict the transition from shear- to buoyancy-driven turbulence,
in terms of turbulent kinetic energy production, energy spectra scaling and mixing layer thickness. In 2020, we investigated the coupled RTKHI system with a Multiple-Relaxation Time (MRT) Discrete Boltzmann Model (DBM) \cite{Chen2020}. To quantitatively analyse the coupled RTKHI process, we resort to morphological and non-equilibrium analysis techniques. We found that both the total boundary length $L$ of the condensed temperature field and the mean heat flux strength $D_{3,1}$ can be used to quantitatively judge the main mechanism in the early stage of the RTKHI system. For the case where the KHI dominates at earlier time and the RTI dominates at later time, the ending point of linear increasing $L$ or $D_{3,1}$ can work as a geometric or physical criterion for discriminating the two stages.

In this paper, we investigate the effects of initial perturbations on the RTI, KHI and the coupled RTKHI systems with the MRT-DBM. The paper is organized as follows: Section II briefly introduces the methodology of Discrete Boltzmann Modeling (DBM) method \cite{DBM}. Systematic numerical simulations on the effects of initial perturbations are shown and analyzed in Section III. A brief conclusion is given in Section IV.

\section{Discrete Boltzmann modeling method}\label{Methodology}

The DBM\cite{xu2012,xu2015aps,xu2016,Xu2018-Chapter2,Xu2021aas,Xu2021cjcp,Xu2021aaas} is developed from a hybrid of Lattice Boltzmann method (LBM) \cite{ss2001,ss1992,ShanChen1,Qin2005PRE,SofoneaSpringer,SXW2011,li2012,Qian2020,Shu2018,Tian2011,Tian2018,Liangpre2014,Zhong2020,Cbx2020pre,QiuRF2020PoF,QiuRF2021PRE,SunDK-AML,SunDK-HMT,Chai-AMC} and the phase space description method of statistical physics.
It is one of the concrete applications of statistical physics coarse-grained modeling method in the field of fluid mechanics, and is a further development of the description method of statistical physical phase space in the form of discrete Boltzmann equation.
Its idea originated from a research review published by Xu et al. in 2012 \cite{xu2012}. In the process of development, it was inspired by the morphological phase space description method \cite{Xu2021cjcp,Xu2021aaas,Xu2009,Xu2010sc,Xu2016sc}.
The methodology of DBM is as follows: It selects a perspective to study a set of kinetic properties of the system according to research requirements. Therefore, the kinetic moments describing this set of properties are required to maintain their values in the model simplification. Based on the independent components of the non-conserved kinetic moments of $(f-f^{eq})$, construct the phase space, and the phase space and its subspaces are used to describe the non-equilibrium behaviors of the system. The research perspective and modeling accuracy should be adjusted as the research progresses, where $f^{eq}$ is the corresponding equilibrium distribution function of $f$.

The establishment of a DBM model needs to go through three steps. The first step is the linearization of the collision term, and the second step is the discretization of the particle velocity space. The principle for coarse-grained modeling is that the physical quantities we choose to measure the system must keep the same values after simplification. The third step is the purpose and core of DBM modeling, and the specific scheme of non-equilibrium state and behavior description should be given. DBM is mainly aimed at the ``mesoscale" and ``dilemma" situations where continuum modeling fails or physical functions are insufficient and molecular dynamics method is unable to do due to the limited applicable scale. The physical information provided by DBM is between macroscopic continuous description and microscopic molecular dynamics. Compared with macroscopic description, DBM observes the system from a wider perspective. The necessity and benefit of non-conserved moment description in DBM increase with the increase of non-equilibrium degree.

Generally, different from the traditional LBM, whose main function is to recover macroscopic fluid equations, the DBM does not aim to solve the macro fluid equations, but mainly to describe and measure both hydrodynamic and thermodynamic nonequilibrium effects beyond the macro fluid equations. DBM only puts forward a physical requirement for the used discrete velocities and does not include specific discrete format. The physical requirement is that the kinetic moments concerned must keep the value unchanged after being converted into sums for calculation. DBM does not use the ``lattice gas" image of ``virtual particle propagation $+$ collision". This is also one reason why it should no longer be called the ``lattice" Boltzmann method. A more detailed introduction can be seen in our recent review articles \cite{Xu2021aas,Xu2021cjcp,Xu2021aaas}.

The DBM equation with external force term can be written as :
\begin{equation}
\frac{\partial f_{i}}{\partial t}+v_{i\alpha }\frac{\partial f_{i}}{\partial x_{\alpha }} = -M_{il}^{-1}[\hat{S}_{lk}(\hat{f}_{k}-\hat{f}_{k}^{eq})+\hat{A}_{l}]-g_{\alpha }\frac{(v_{i\alpha }-u_{\alpha})}{RT}f_{i}^{eq}\text{,}
\end{equation}%
where the variable $t$ and $x_{\alpha}$ are the time and spatial coordinates, $T$ is the temperature, $g_{\alpha}$ and $u_{\alpha}$ denote the acceleration and velocity in the $x_{\alpha}$ direction, $v_{i\alpha}$ is the discrete particle velocity, $i=1$, $\ldots$, $N$, and the subscript $\alpha$ indicates the $x$, $y$, or $z$ component. $f_{i}$ and $\hat{f}_{i}$ ($f_{i}^{eq}$ and $\hat{f}_{i}^{eq}$) are the particle (equilibrium) distribution functions in velocity space and kinetic moment space, respectively; the mapping between moment space and velocity space is defined by the linear transformation $M_{ij}$, i.e., $\hat{f}_{i}=M_{ij}f_{j}$ and $f_{i}=M_{ij}^{-1}\hat{f}_{j}$. The matrix $\hat{S}={\rm diag} (s_{1},s_{2},\cdots ,s_{N})$ is the diagonal relaxation matrix. $\hat{A}_{l}$ is the correction term used for recovering reasonable hydrodynamic behaviors.

In the paper, the following two-dimensional discrete velocity model is used:
\begin{align}
\left(v_{ix,}v_{iy}\right) =\left\{
\begin{array}{cc}
\mathbf{cyc}\!:c\left( \pm 1,0\right) , & \text{for }1\leq i\leq 4, \\
c\left( \pm 1,\pm 1\right) , & \text{for }5\leq i\leq 8, \\
\mathbf{cyc}\!:2c\left( \pm 1,0\right) , & \text{for }9\leq i\leq 12, \\
2c\left( \pm 1,\pm 1\right) , & \text{for }13\leq i\leq 16,%
\end{array}%
\right.  \label{dvm3}
\end{align}%
and $\eta_{i}=\eta_{0}$ for $i=1$, \ldots, $4$, and $\eta_{i}=0$ for $i=5 $, \ldots , $16$, which is introduced to control the specific-heat-ratio $\gamma$. ``\textbf{cyc}" indicates the cyclic permutation, and $c$ and $\eta_{0}$ are two free parameters, which are adjusted to optimize the properties of the model. The specific forms of the correction term, transformation matrix and the corresponding equilibrium distribution functions in kinetic moment space can be seen in the appendix for details.

The kinetic moments of $f_{i}-f_{i}^{eq}$ can be used to check and measure the Thermodynamic Non-Equilibrium (TNE) state and effects. $\Delta_{i}^{\ast}=M_{ij}^{\ast }(f_{j}-f_{j}^{eq})$,
where $M_{ij}^{\ast}(f_{i})$ represents kinetic central moments of $f_{i}$, and its specific expression can be obtained by replacing the variable $v_{i\alpha}$ by $v_{i\alpha}-u_{\alpha}$ in $M_{ij}$. At present, the commonly used non-equilibrium characteristic quantities are $\Delta_{2\alpha \beta }^{\ast}$,  $\Delta_{(3,1)\alpha}^{\ast}$, $\Delta_{3\alpha \beta \gamma}^{\ast}$ and $\Delta_{(4,2)\alpha \beta}^{\ast}$. $\Delta_{2\alpha \beta }^{\ast}$ indicates the viscous stress tensor and $\Delta_{(3,1)\alpha}^{\ast}$ indicates the heat flux. $\Delta_{3\alpha \beta \gamma}^{\ast}$ and $\Delta_{(4,2)\alpha \beta}^{\ast}$ indicate the flux of viscous stress and of heat flux, respectively. Where, the subscripts ``2" and ``3" indicate tensors of second order and third order, the subscript ``3,1" means a first-order tensor contracted from a third-order tensor, and the subscript ``4,2" means a second-order tensor contracted from a fourth-order tensor.

In the phase space opened by the non-conserved kinetic moments and their subspaces, the origin of coordinates corresponds to a thermodynamic equilibrium state, and any other point corresponds to a specific thermodynamic non-equilibrium state. The corresponding non-equilibrium strength can be defined by means of the distance from the origin. For example, the mean heat flux strength can be defined as $D_{(3,1)}=\overline{\sqrt{\Delta_{(3,1)\alpha}^{\ast 2}}}$. These more abundant but previously poorly understood characteristics of non-equilibrium behavior contain a large number of physical functions to be explored. At the same time, it should be pointed out that in addition to the idea of DBM modeling, a series of data processing techniques, characteristic scale extraction technology and structure analysis techniques developed in DBM application process can be directly used for data processing and feature analysis of various complex configuration and dynamic physical field.

\section{Numerical Simulations}

In recent decades, numerous investigations have shown that the initial conditions have important effects on the evolution of interface instability. In this section, we investigate the effects of initial perturbations on the RTI, KHI and the coupled RTKHI systems with the multiple-relaxation-time DBM model, which has been validated by some well-known benchmark tests\cite{Chen2016, Chen2018, Chen2020}.

%%%%%%%%%%%%%%%%%%%%%%%%%%%%%%%%%%%%%%%%
\begin{figure}[tbp]
\center\includegraphics*%
%[bbllx=30pt,bblly=15pt,bburx=240pt,bbury=220pt,width=0.4\textwidth]{Fig1.pdf}
[width=0.4\textwidth]{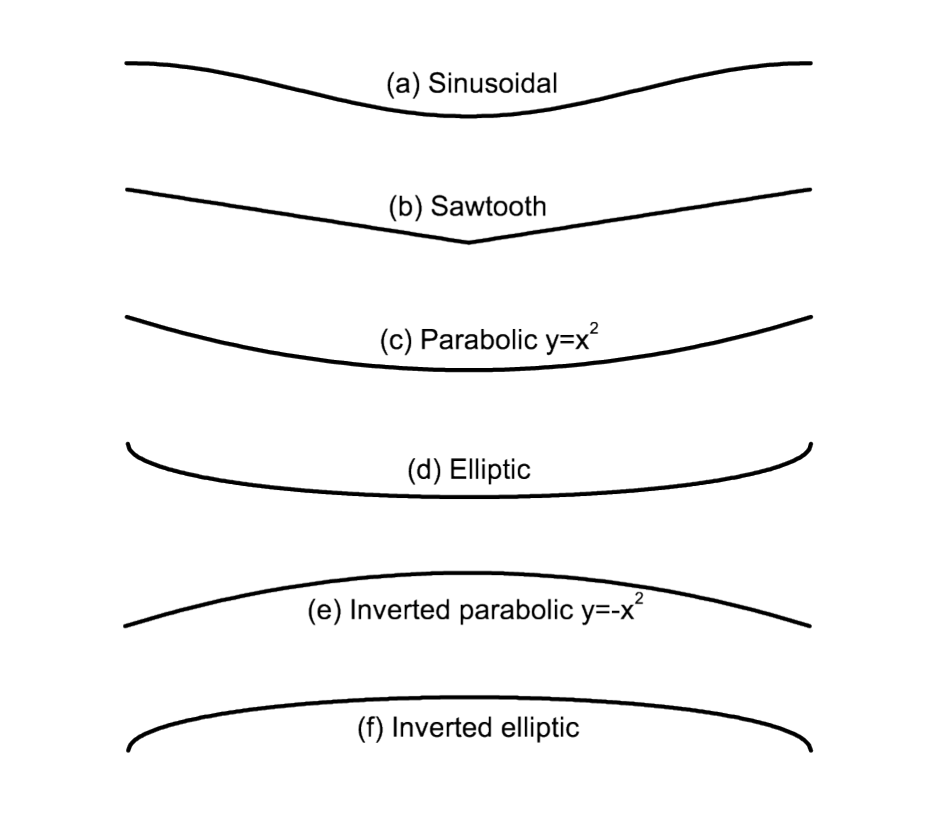}
\caption{The schematic diagram of the initial disturbance interface. (a) sinusoidal perturbation, (b) sawtooth perturbation, (c) parabolic $y=x^{2}$, (d) elliptic disturbance, (e) inverted parabolic $y=-x^{2}$, and (f) inverted elliptic disturbance interface.}
\end{figure}
%%%%%%%%%%%%%%%%%%%%%%%%%%%%%%%%%%%%%%%%%
%%%%%%%%%%%%%%%%%%%%%%%%%%%%%%%%%%%%%%%%
\begin{figure*}
\begin{center}
\includegraphics[width=0.7\textwidth]{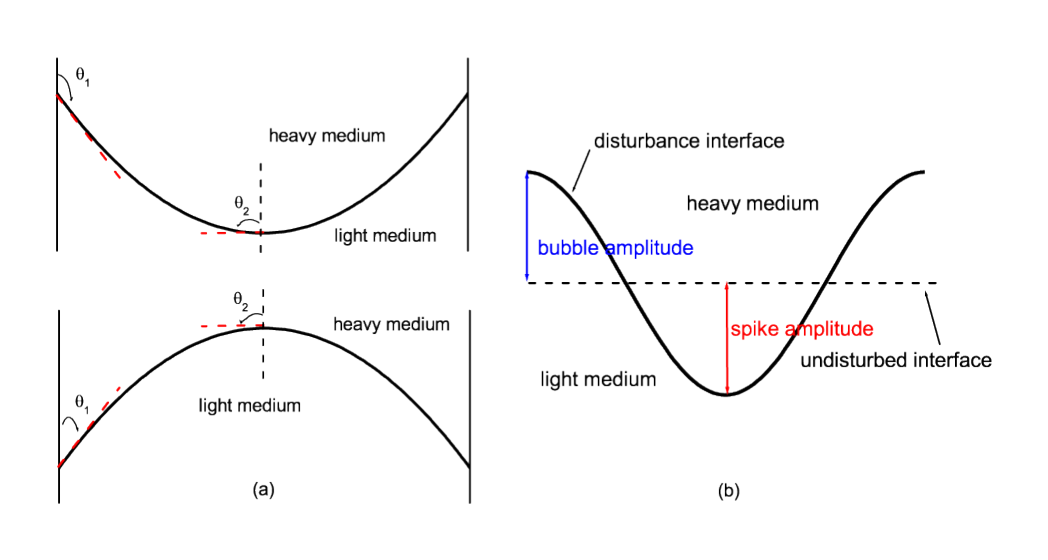}
\end{center}
\caption{Schematic diagram. (a) Definition of the contact angle, $\theta_{1}$ is the bilateral contact angle, $\theta_{2}$ is the middle contact angle. (b) Definition of the bubble amplitude and the spike amplitude.}
\end{figure*}
%%%%%%%%%%%%%%%%%%%%%%%%%%%%%%%%%%%%%%%%%

Six different initial conditions, shown in Fig. 1, are designed to study the effects of the initial perturbation shape on the instability system, including (a) sinusoidal perturbation, (b) sawtooth perturbation, (c) parabolic $y=x^{2}$, (d) elliptic disturbance, (e) inverted parabolic $y=-x^{2}$, and (f) inverted elliptic disturbance interface.
The angle between the tangent at the end of the disturbance interface and the boundary is called the bilateral contact angle $\theta_{1}$, and the angle between the centerline and the tangent at the center point is called the middle contact angle $\theta_{2}$, shown in Fig. 2(a). The two contact angles together describe the basic morphology of the disturbance interface. Except for the sawtooth perturbation, the middle contact angle $\theta_{2}$ of the other five interfaces is approximately equal to $90^{o}$. These six interfaces basically include the continuous curve shape from the boundary to the center line determined by $\theta_{1}$ and $\theta_{2}$. Their related understanding and recognition can be helpful to analyze the multimode disturbance in the future.

In the simulation, a two-dimensional computational domain with height $H=80$ and width $W=20$ is adopted, and divided into $NX \times NY = 101\times400$ mesh-cells. The computational domain can be divided into upper and lower parts by the disturbed interface. In the two parts, two different homogeneous temperatures are fixed, and the corresponding hydrostatic density profiles $\rho(y)$ follow the regularity, $\partial _{y}p(y)=-g_{y}\rho (y)$, and $p(y)=T \rho (y)$ in each part. With fixed $T$, the solution has an exponentially decaying behavior in the two half volumes. The initial hydrostatic unstable configuration is therefore given by
\begin{eqnarray}
&T(y)=T_{u},\;\rho(y)=\rho _{u}\exp(-g_{y}(y-y_{s})/T_{u}), \notag \\
&u_{x}(y)={u_{0}}, \;u_{y}(y)=0, & \text{for } y\geq y_{s}, \notag
\\
&T(y)=T_{b},\;\rho(y)=\rho _{b}\exp (-g_{y}(y-y_{s})/T_{b}), \notag \\
&\;u_{x}(y)=-u_{0},\;u_{y}(y)=0, & \text{for } y < y_{s}, %
\label{rt}%
\end{eqnarray}%
where $\rho_{b}$ and $T_{b}$ ($\rho_{u}$ and $T_{u}$) are the density and temperature of the light (heavy ) fluid, $u_{0}$ is the shear velocity, $g_{y}$ is the gravitational acceleration, $y_{s}$ is the initial disturbance interface. For these six interfaces, the specific expressions of initial disturbances are as follows:
\begin{subequations}
\begin{equation}
y_{s}^{a}=H/2+a_{0}\cos (2\pi x/\lambda),
\end{equation}%
\begin{equation}
y_{s}^{b}=H/2-a_{0}+2a_{0}\frac{\vert{ x-\lambda /2 \vert}}{\lambda /2},
\end{equation}%
\begin{equation}
y_{s}^{c}=H/2-a_{0}+2a_{0}\frac{(x-\lambda /2)^{2}}{(\lambda /2)^{2}},
\end{equation}%
\begin{equation}
y_{s}^{d}=H/2+a_{0}-2a_{0}(1-\frac{(x-\lambda /2)^{2}}{(\lambda /2)^{2}})^{0.5},
\end{equation}%
\begin{equation}
y_{s}^{e}=H/2+a_{0}-2a_{0}\frac{(x-\lambda /2)^{2}}{(\lambda /2)^{2}},
\end{equation}%
\begin{equation}
y_{s}^{f}=H/2-a_{0}+2a_{0}(1-\frac{(x-\lambda /2)^{2}}{(\lambda /2)^{2}})^{0.5},
\end{equation}%
\end{subequations}
where $\lambda=W$ is the wavelength of the perturbation, $a_{0}$ is the initial amplitude. To have a finite width of the initial interface, a smooth interpolation between the two half-volumes is plused on the initial configuration. When the relative velocity $u_{0}$ is set to zero, this is a pure RTI system; when the acceleration $g_{y}$ is set to zero, it is a pure KHI; with various tangential velocities and accelerations, the coupled RTKHI systems are gotten. The fifth-order weighted essentially non-oscillatory (WENO) scheme is adopted for space discretization, and the third-order Runge-Kutta scheme is adopted for time discretization. The left and right boundaries are periodic boundary conditions. In the bottom and the top boundaries, we apply the solid wall boundary conditions, i.e.,
$(\rho, U_{x}, T)|_{ix,-2}=(\rho, U_{x}, T)|_{ix,-1}=(\rho, U_{x}, T)|_{ix,0}=(\rho, U_{x}, T)|_{ix,1}$, $U_{y}|_{ix,-2}=U_{y}|_{ix,-1}=U_{y}|_{ix,0}=0$,
where $(ix,-2)$, $(ix,-1)$, and $(ix,0)$ are the indexes of ghost nodes out of the bottom boundary $(ix,1)$ when the fifth-order WENO scheme is used. The distribution functions $f_{i}$ on the ghost nodes are approximated by theirs equilibrium distribution functions $f_{i}^{eq}$. On the top side, we can operate in a similar way.

We first investigate the influence of the initial interface shape on the evolution of RTI. Figure 3 shows the temperature patterns of the above six different initial disturbances evolving to $t=300$ under the action of RTI. Figure 4 shows the time evolutions of bubbles, spikes, disturbance amplitudes and morphological boundary lengths for different initial interface systems. The definitions of bubble and spike amplitudes are shown in Fig. 2(b). The disturbance amplitude of RTI is defined as half of the sum of the bubble amplitude and the spike amplitude. The morphological boundary length $L$ is the total length of the interface between high and low temperature (light and heavy) fluids. The methodology of morphological analysis technique is as follows: By defining a temperature threshold $T_{th}$, the value higher than the threshold is defined as the high temperature area (white area), and the value below the threshold is defined as the low temperature area (black area). In this way, the original continuous temperature field is transformed into a temperature pattern of black and white pixels, the total length of the dividing lines between white and black regions is defined as the morphological boundary length $L$. The specific calculation method of $L$ can refer to our previous work \cite{Xu2021cjcp,Xu2009,Xu2010sc,Chen2020}. Both disturbance amplitude and morphological boundary length are important parameters that describe the degree of instability development and material mixing. They have different perspectives, but they are related and cannot replace each other. In the simulation, the dimensionless parameters are set as follows: $\rho_{b}=1$, $T_{b}=1.4$, $\rho_{u}=2.33333$, $T_{u}=0.6$, $g_{x}=0$, $g_{y}=0.005$, $u_{0}=0$, $c=1$, $\eta_{0}=3$, temperature threshold $T_{th}=1.0$, the specific-heat-ratio $\gamma=1.4$, the initial amplitude $a_{0}=2$, and spatial step $dx=dy=0.2$, temporal step $dt=10^{-3}$.

It can be seen from Figs. 3 and 4 that the initial disturbance interface shape has a great influence on the evolution of RTI. In the evolution of RTI, the heavy and light fluids gradually penetrate into each other as time progresses, with the light fluid rising to form a bubble and the heavy fluid falling to generate a spike. The sharper the interface (the interface has a higher curvature), the faster the growth of bubble or spike. For example, in the system with an inverted elliptic interface, the heavy fluid falls along both sides of the computing domain to form spikes. Compared with other interfaces where spikes are formed, the interface in this place is the sharpest, and the spike falls at the fastest rate; The light fluid rises in the middle part of the interface to form bubbles, compared with other interfaces to form bubbles, the interface here is the bluntest and the bubble rises at the slowest speed. The difference of amplitudes of RTI systems with different initial interfaces is small (Fig. 4(c)), but the morphological boundary length differs greatly in the later stage (Fig. 4(d)), indicating that the initial interface shape has substantial influence on the mixing degree of different media in the RTI system.

%%%%%%%%%%%%%%%%%%%%%%%%%%%%%%%%%%%%%%%%
\begin{figure}[tbp]
\center\includegraphics*%
%[bbllx=5pt,bblly=125pt,bburx=585pt,bbury=510pt,width=0.5\textwidth]{Fig3.pdf}
[width=0.5\textwidth]{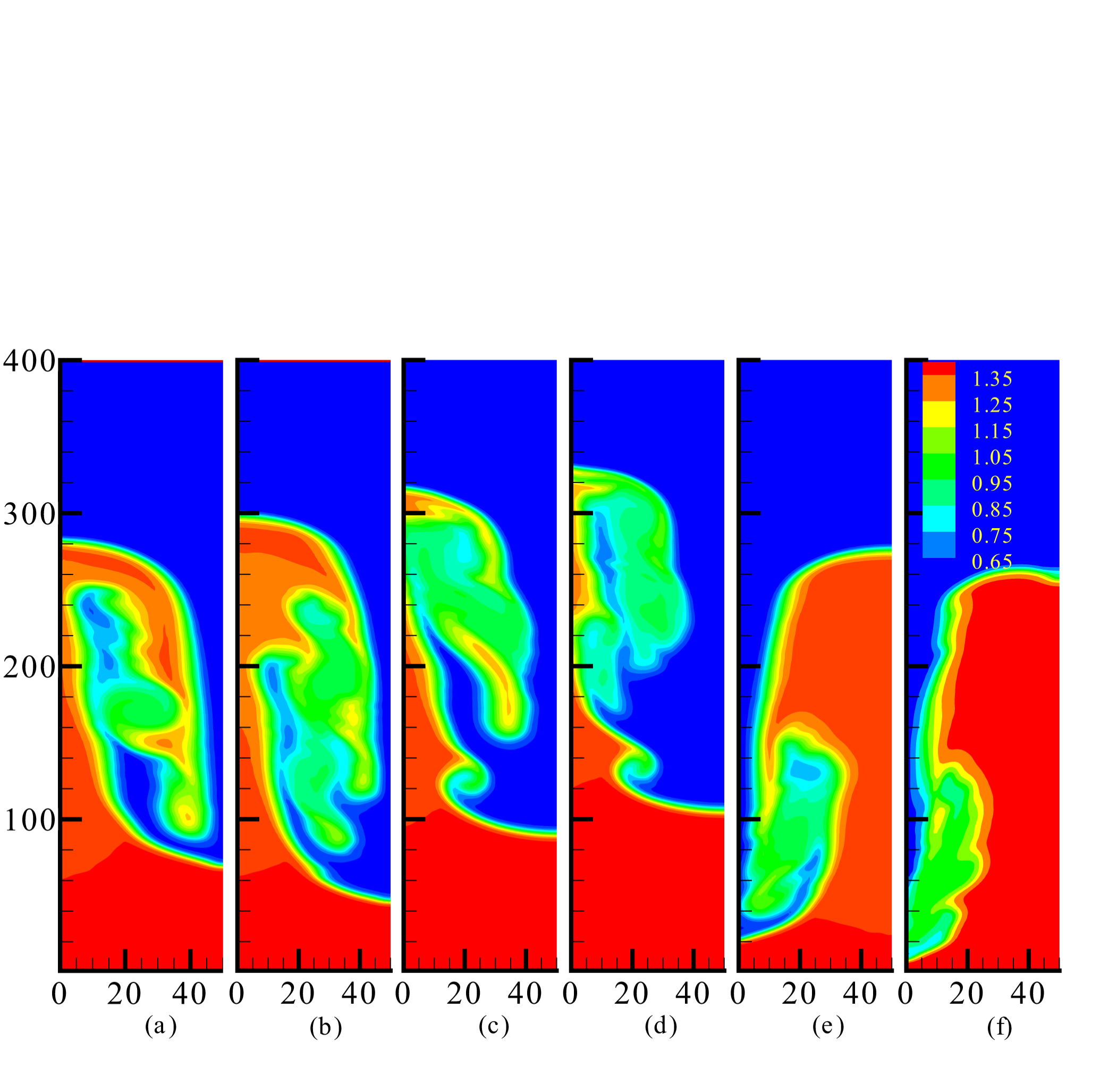}
\caption{(Color online) The influence of the initial interface shape on the evolution of RTI. (a) to (f) are the temperature patterns of the six systems with different initial disturbances described in Fig. 1 at time $t=300$. The horizontal and vertical coordinates are the mesh numbers in the computing domain. For clarity, only half of the computational domains are shown. From blue to red corresponds to an increase of temperature. Each figure follows the same legend.}
\end{figure}
%%%%%%%%%%%%%%%%%%%%%%%%%%%%%%%%%%%%%%%%%

%%%%%%%%%%%%%%%%%%%%%%%%%%%%%%%%%%%%%%%%
\begin{figure}[tbp]
\center\includegraphics*%
%[bbllx=15pt,bblly=5pt,bburx=295pt,bbury=225pt,width=0.5\textwidth]{Fig4.pdf}
[width=0.5\textwidth]{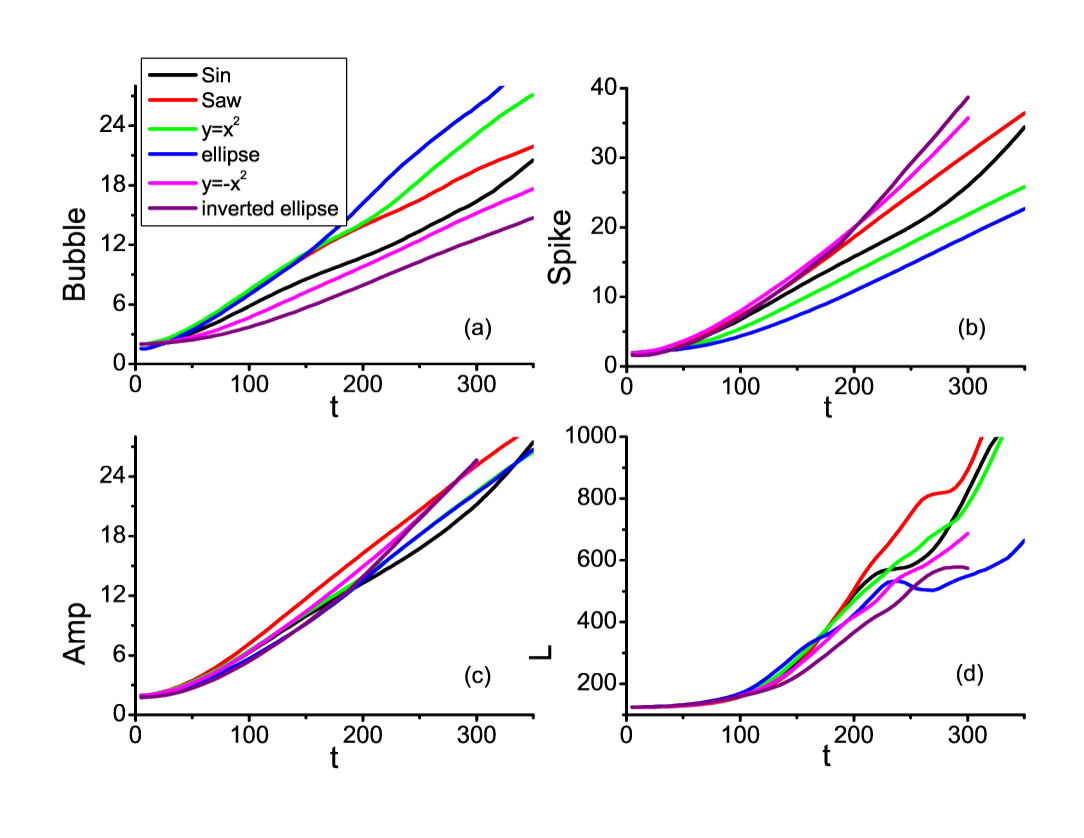}
\caption{(Color online) Effects of initial interface shape on RTI evolution, (a) bubble amplitude, (b) spike amplitude, (c) total disturbance amplitude (= half of the sum of bubble amplitude and spike amplitude), (d) morphological boundary length.}
\end{figure}
%%%%%%%%%%%%%%%%%%%%%%%%%%%%%%%%%%%%%%%%%

%%%%%%%%%%%%%%%%%%%%%%%%%%%%%%%%%%%%%%%%
\begin{figure}[tbp]
\center\includegraphics*%
%[bbllx=5pt,bblly=250pt,bburx=595pt,bbury=410pt,width=0.5\textwidth]{Fig5.pdf}
[width=0.5\textwidth]{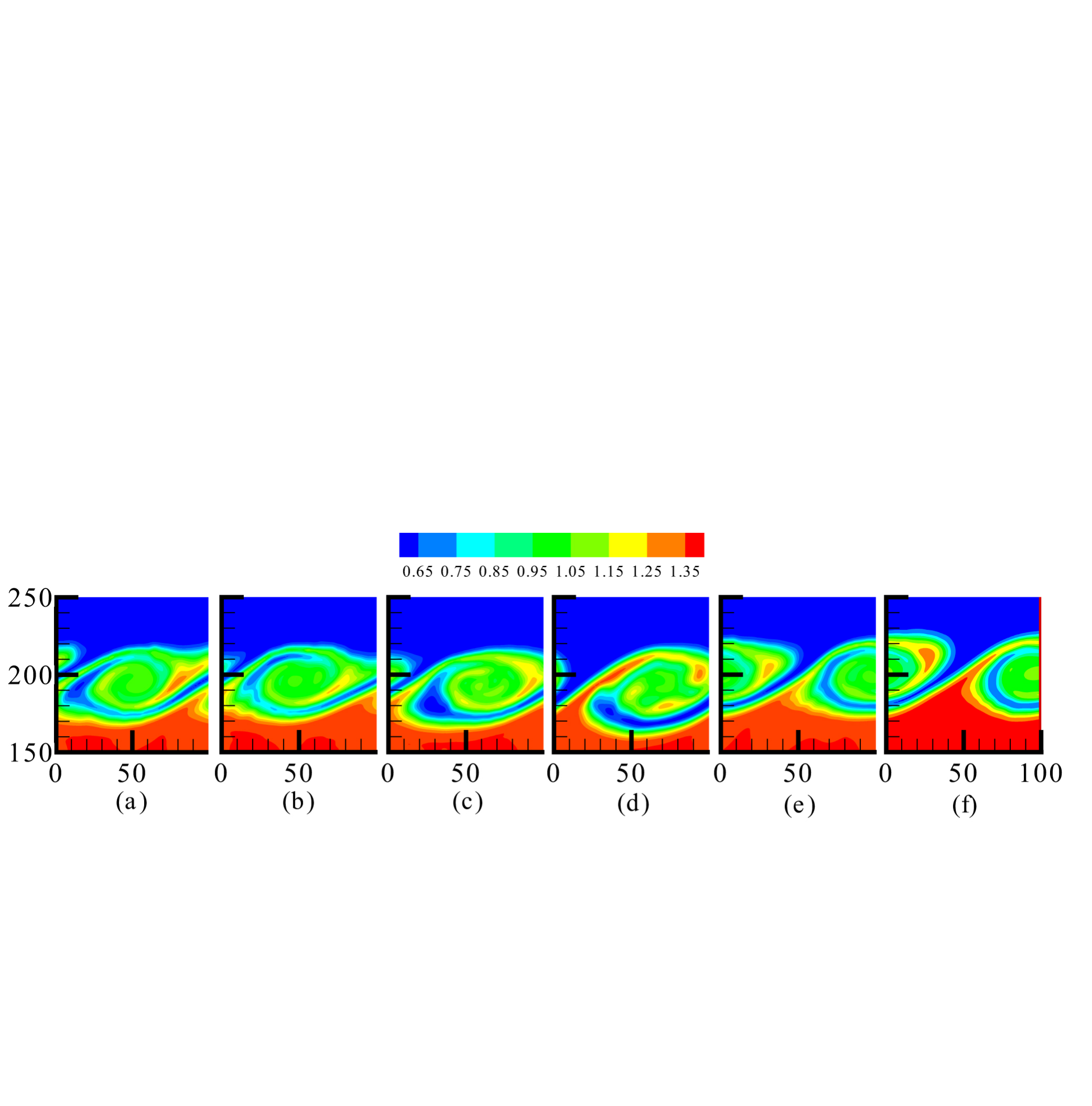}
\caption{(Color online) The influence of the initial interface shape on the evolution of KHI. (a) to (f) are the temperature patterns of the six systems with different initial disturbances described in Fig. 1 at time $t=300$. From blue to red corresponds to an increase of temperature. Each figure follows the same legend.}
\end{figure}
%%%%%%%%%%%%%%%%%%%%%%%%%%%%%%%%%%%%%%%%%

%%%%%%%%%%%%%%%%%%%%%%%%%%%%%%%%%%%%%%%%
\begin{figure}[tbp]
\center\includegraphics*%
%[bbllx=15pt,bblly=15pt,bburx=300pt,bbury=120pt,width=0.5\textwidth]{Fig6.pdf}
[width=0.5\textwidth]{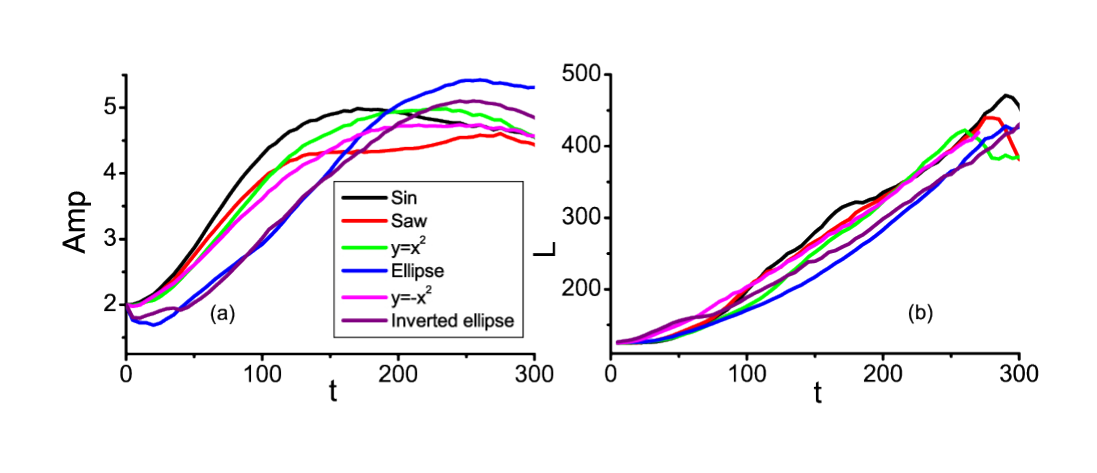}
\caption{(Color online) Effects of initial interface shape on KHI evolution, (a) amplitude (1/2 width of the mixing layer), (b) morphological boundary length $L$ of the temperature field.}
\end{figure}
%%%%%%%%%%%%%%%%%%%%%%%%%%%%%%%%%%%%%%%%%

Figure 5 shows the temperature patterns of the six different initial disturbances evolving to $t=300$ under the action of KHI. Figure 6 shows the time evolutions of amplitudes and morphological boundary lengths for different KHI systems. The amplitude of KHI systems is defined as half of the mixing width. The initial parameters are $g_{x}=0$, $g_{y}=0$, $u_{0}=0.1$, and the other parameters are the same as Figure 3. It can be seen from Figs. 5 and 6 that the initial interface shape has little influence on the evolution of KHI. There is no significant difference in the final degree of material mixing in KHI systems with different initial interface shapes.

%%%%%%%%%%%%%%%%%%%%%%%%%%%%%%%%%%%%%%%%
\begin{figure}[tbp]
\center {
{\epsfig{file=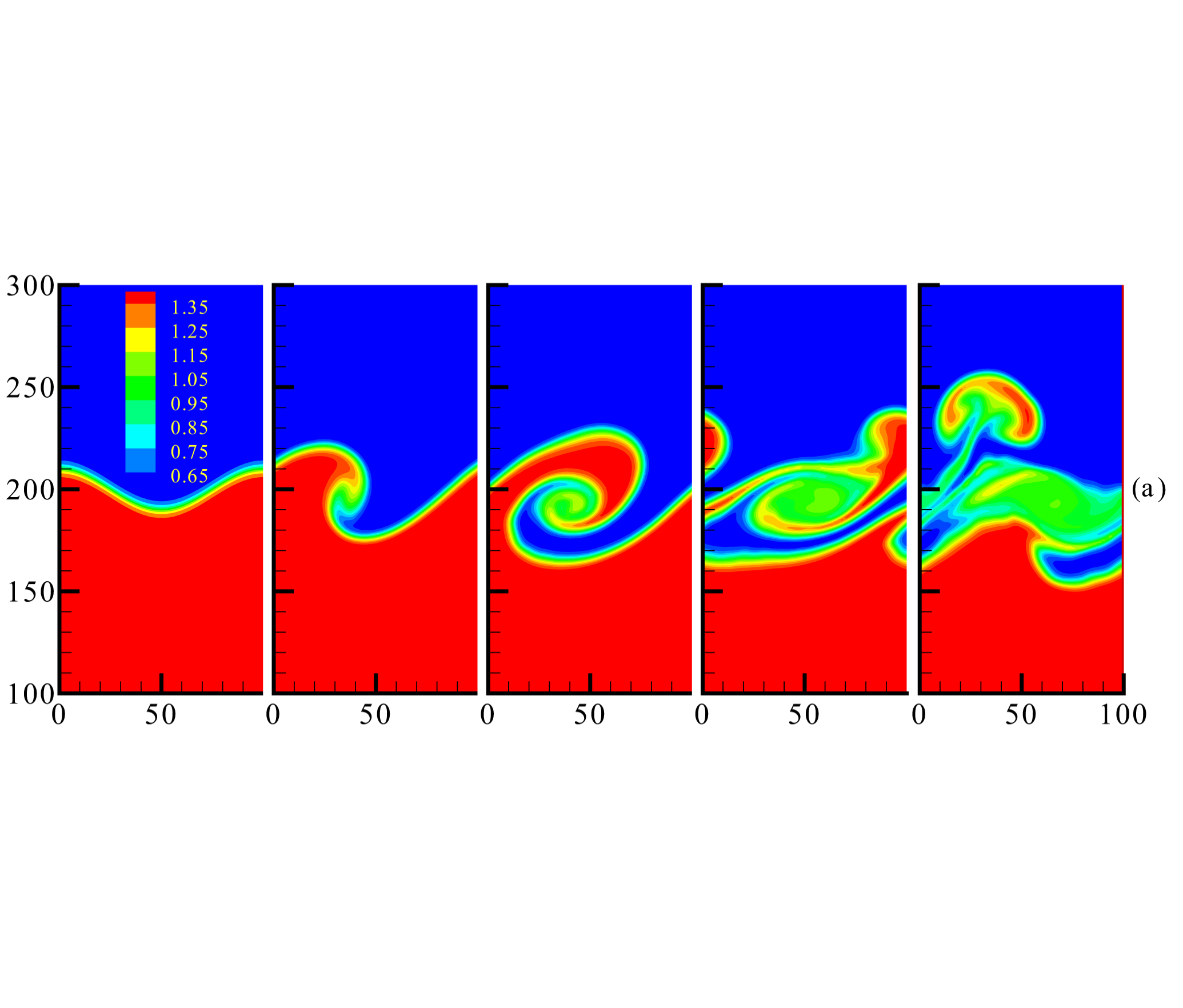,%bbllx=5pt,bblly=280pt,bburx=598pt,bbury=515pt,
%height=0.19\textwidth,width=0.48\textwidth,clip=}}\\
height=0.19\textwidth,
width=0.48\textwidth}}\\
{\epsfig{file=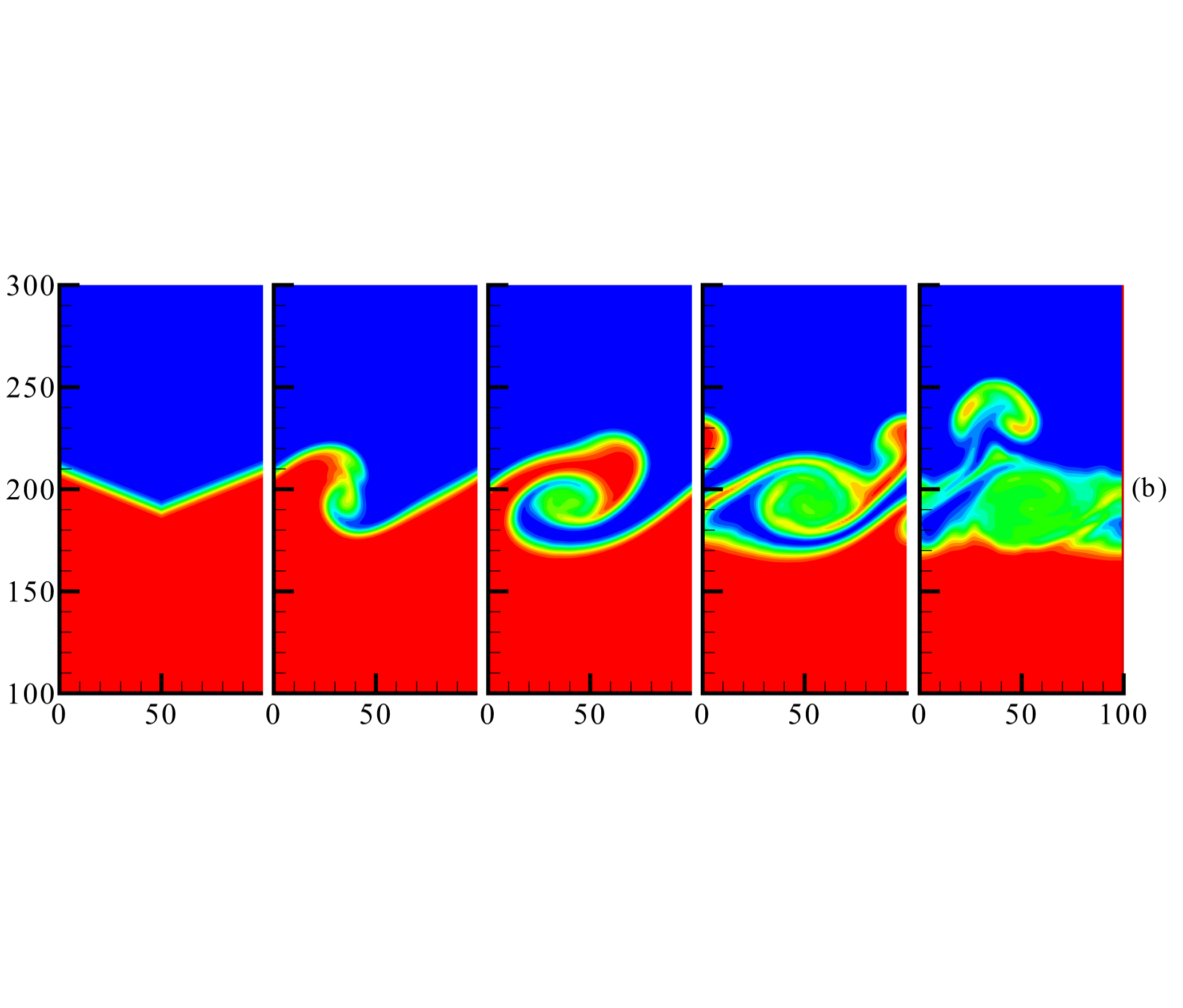,%bbllx=5pt,bblly=280pt,bburx=598pt,bbury=515pt,
%height=0.19\textwidth,width=0.48\textwidth,clip=}}\\
height=0.19\textwidth,
width=0.48\textwidth}}\\
{\epsfig{file=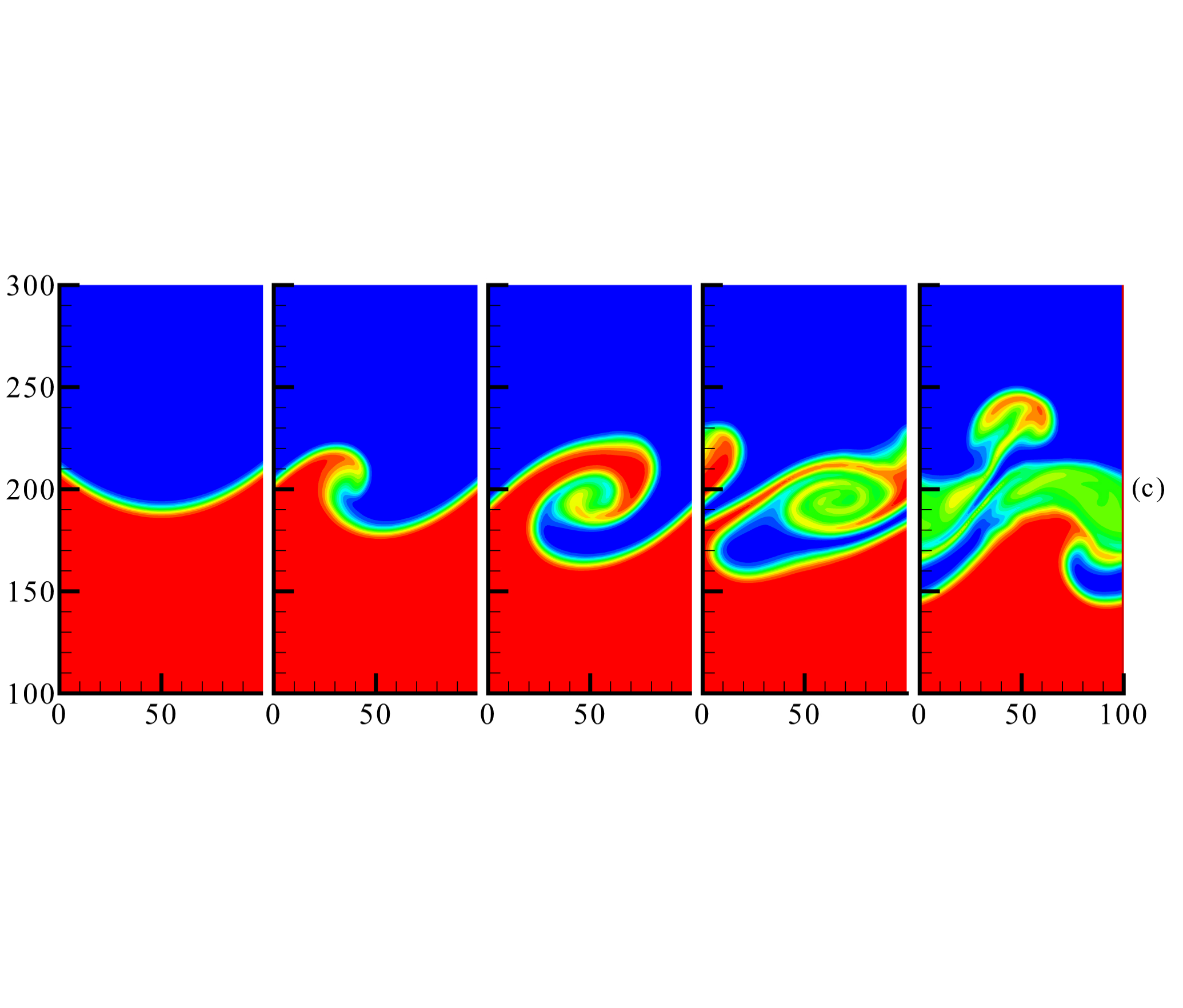,%bbllx=5pt,bblly=280pt,bburx=598pt,bbury=515pt,
%height=0.19\textwidth,width=0.48\textwidth,clip=}}\\
height=0.19\textwidth,
width=0.48\textwidth}}\\
{\epsfig{file=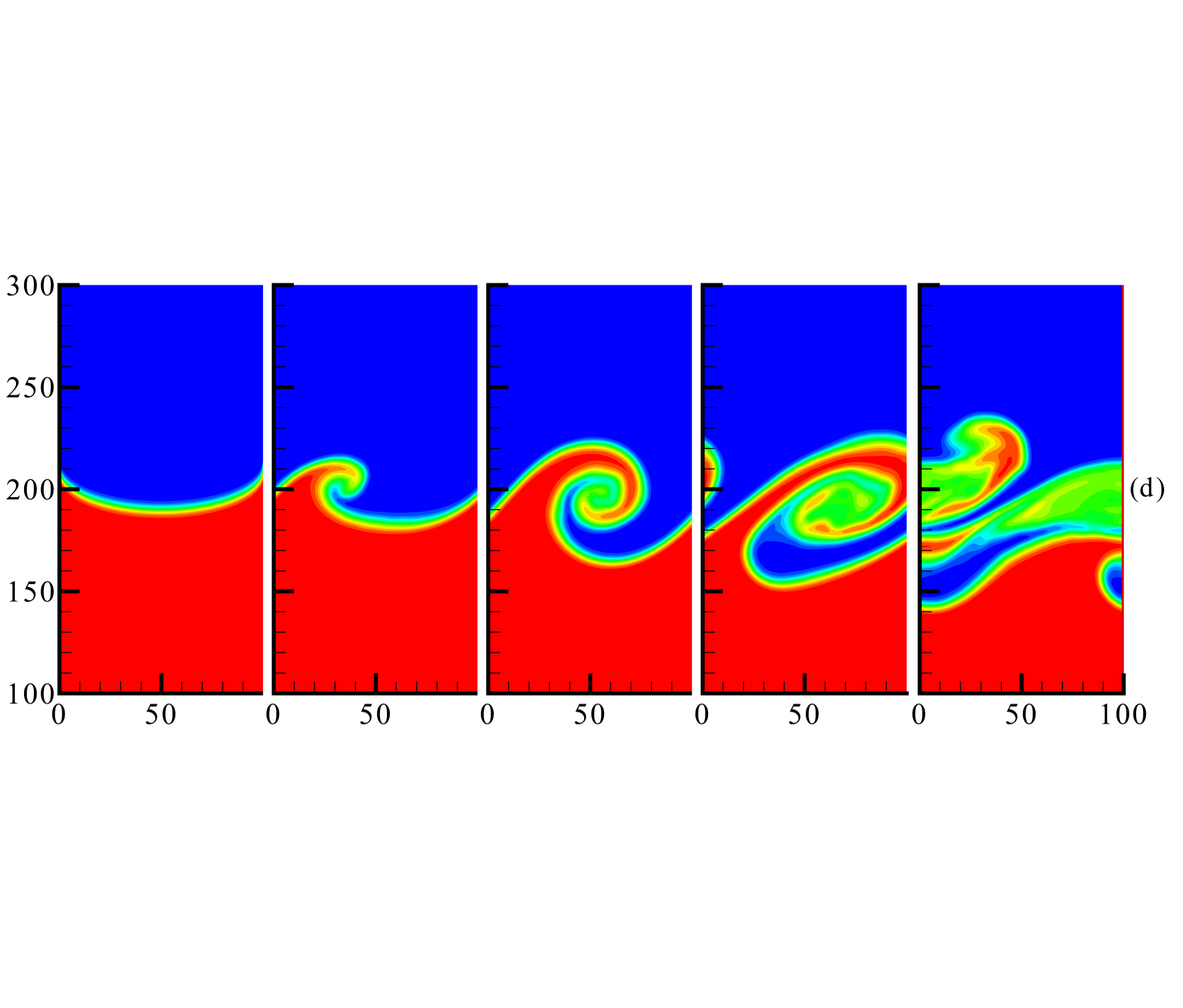,%bbllx=5pt,bblly=280pt,bburx=598pt,bbury=515pt,
%height=0.19\textwidth,width=0.48\textwidth,clip=}}\\
height=0.19\textwidth,
width=0.48\textwidth}}\\
{\epsfig{file=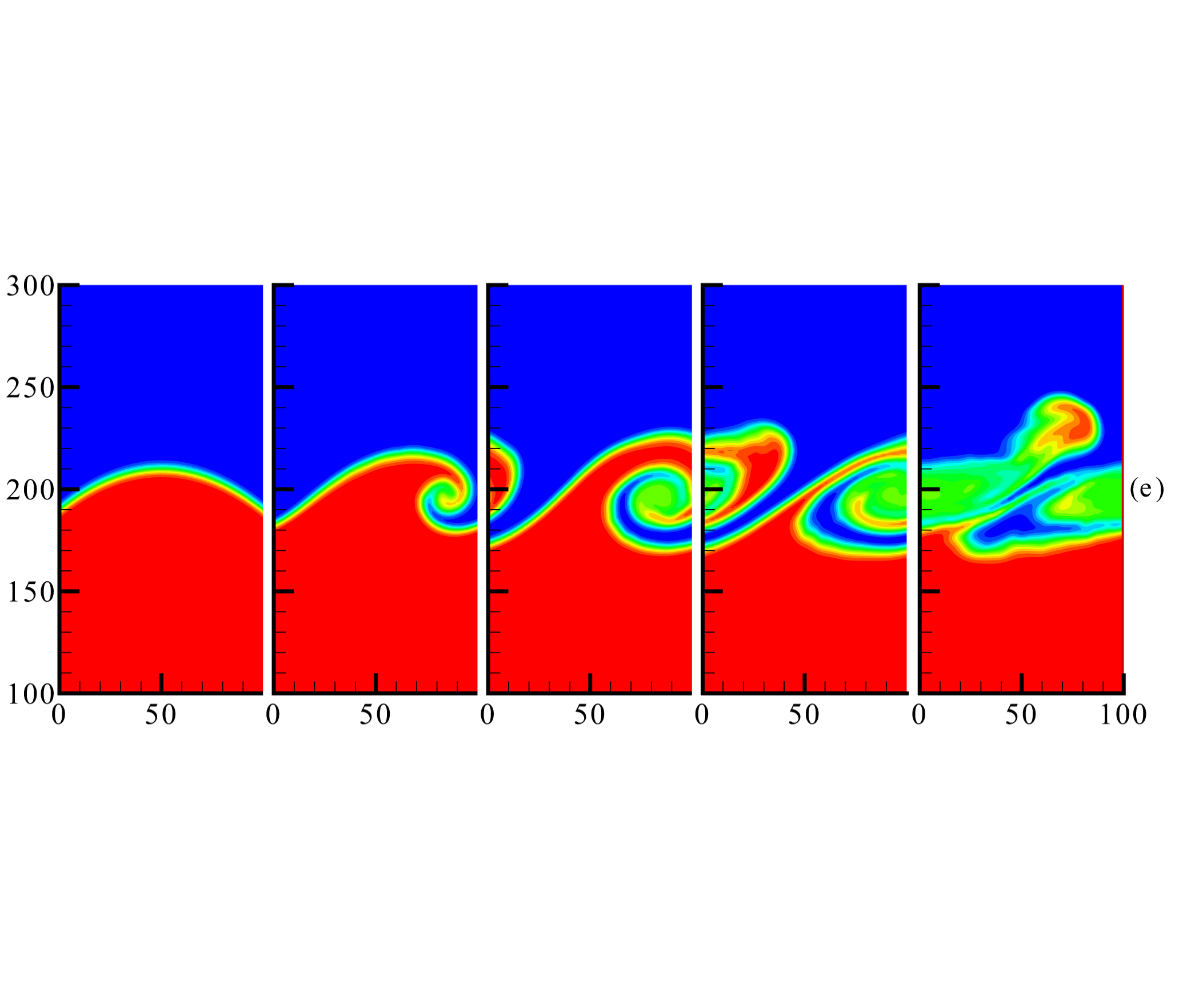,%bbllx=5pt,bblly=280pt,bburx=598pt,bbury=515pt,
%height=0.19\textwidth,width=0.48\textwidth,clip=}}\\
height=0.19\textwidth,
width=0.48\textwidth}}\\
{\epsfig{file=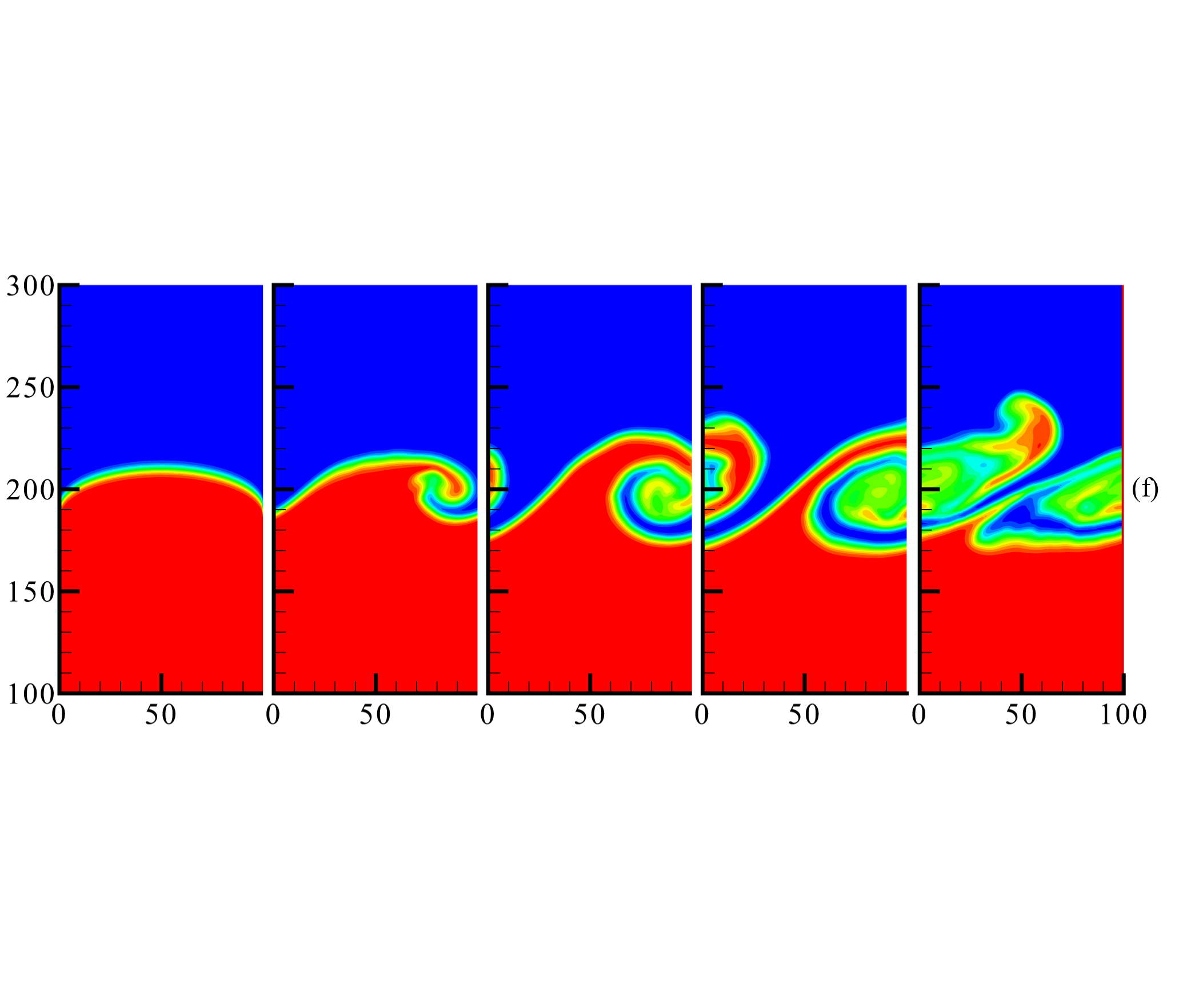,%bbllx=5pt,bblly=280pt,bburx=598pt,bbury=515pt,
%height=0.19\textwidth,width=0.48\textwidth,clip=}}
height=0.19\textwidth,
width=0.48\textwidth}}
}
\caption{(Color online) Temperature patterns of the RTKHI. From (a) to (f), the initial perturbations are sinusoidal, sawtooth, parabolic, ellipse, inverted parabolic and inverted elliptic perturbation interfaces, respectively. The images from left to right correspond to $t=0, 50, 100, 150, 200$, respectively. Each figure follows the same legend.}\label{Fig-RTKHI-T}
\end{figure}
%%%%%%%%%%%%%%%%%%%%%%%%%%%%%%%%%%%%%%%%%

Figure \ref{Fig-RTKHI-T} shows the evolutions of the coupled RTKHI systems with different initial disturbances at times $t=0$, $50$, $100$, $150$, and $200$. The initial conditions are $g_{x}=0$, $g_{y}=0.005$, $u_{0}=0.15$, and the other parameters remain the same. As shown in the figure, under the action of initial perturbation and tangential velocity, the perturbations gradually grow to rolled-up vortexes, which are the main characteristic of free shear flows, so KHI plays a major role at the initial stage. For the inverted parabolic and elliptic interfaces, the vortexes are rolled up on the right side of the computational domain, which is different from the other four interfaces. As time progresses, more fluid is entrained in the vortical structure, the secondary RTI develops along the vortex arms, and plays a major role.
It should be pointed out that in the systems with different disturbance interfaces, the secondary RTI develops at different times and with different speeds. Among them, the first three systems develop relatively fast, while the last three develop relatively slowly. In other words, the shape of the perturbed interface has a certain degree of influence on the evolution of the coupled RTKHI system. The focal points of the following research are: (i) the influence of perturbed interface on the transition point of the coupled system; (ii) the influence of perturbed interface on the main mechanism in the early stage of the RTKHI system.

%%%%%%%%%%%%%%%%%%%%%%%%%%%%%%%%%%%%%%%%
\begin{figure}[tbp]
\center\includegraphics*
%[bbllx=15pt,bblly=15pt,bburx=215pt,bbury=220pt,width=0.50\textwidth]{Fig8.pdf}
[width=0.50\textwidth]{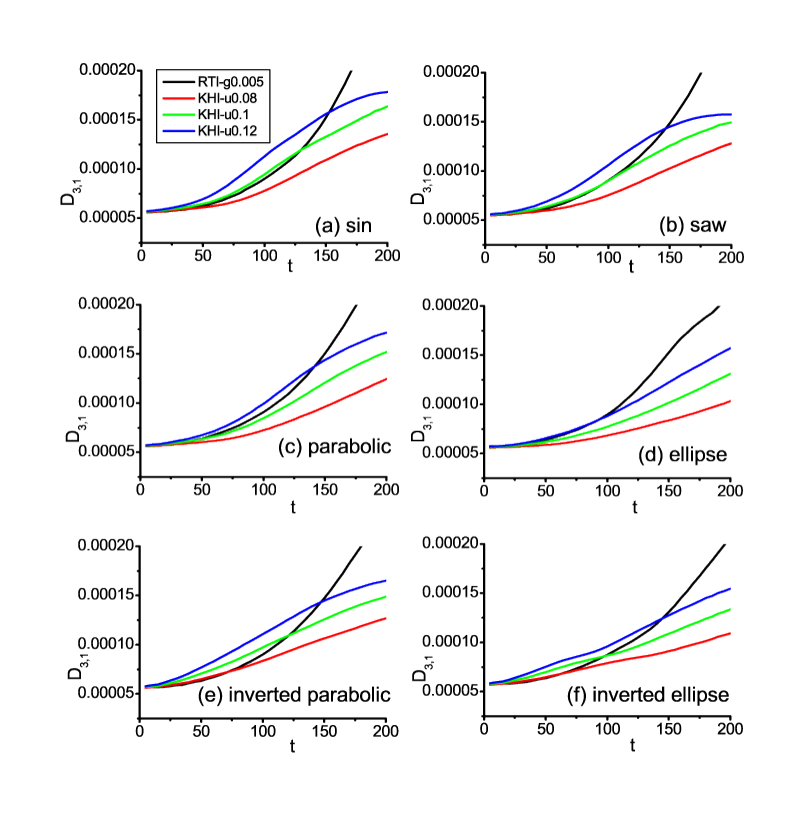}
\caption{(Color online) Influence of the initial interface shapes on the early mechanism judgment of the coupled system.}\label{Fig-sm}
\end{figure}
%%%%%%%%%%%%%%%%%%%%%%%%%%%%%%%%%%%%%%%%%
%%%%%%%%%%%%%%%%%%%%%%%%%%%%%%%%%%%%%%%%
\begin{figure}[tbp]
\center\includegraphics*%
%[bbllx=15pt,bblly=15pt,bburx=250pt,bbury=230pt,width=0.51\textwidth]{Fig9.pdf}
[width=0.51\textwidth]{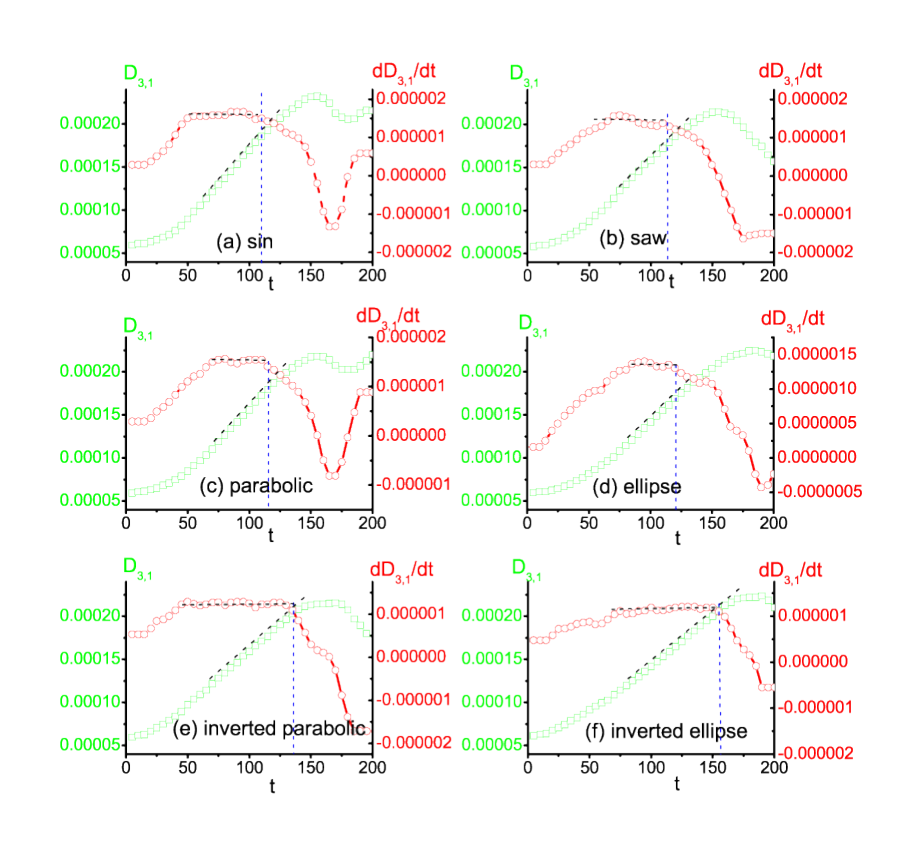}
\caption{(Color online) Influence of the initial interface shapes on the transition points of the coupled system. From (a) to (f), the initial perturbations are sinusoidal, sawtooth, parabolic, ellipse, inverted parabolic, and inverted elliptic perturbation interfaces, respectively.}\label{Fig-sp}
\end{figure}
%%%%%%%%%%%%%%%%%%%%%%%%%%%%%%%%%%%%%%%%%

In the previous work, we have shown that the main mechanism of the coupled RTKHI system in the early stage depends on the comparison of buoyancy and shear strength, i.e., gravity acceleration $g$ and shear velocity $u_ {0}$, and both the morphological boundary length $L$ and the mean heat flux strength $D_{3,1}$ can be used to quantitatively judge the main mechanism in the early stage. The judgment method is as follows:  Comparing the morphological boundary length $L$ or the non-equilibrium strength $D_{3,1}$ of pure RTI with acceleration $g$ and KHI systems with shear velocity $u_{0}$, if $L^{KHI} = L^{RTI}$ ($D_{3,1}^{KHI} = D_{3,1}^{RTI}$), the buoyancy and shear effects are balanced in the early stages of the corresponding coupled system; The shear velocity $u_{0}$ is defined as the critical velocity $u_{C}$ of the coupled system. It stands to reason that, keeping $g$ constant, if the shear velocity increases, the shear effect is stronger than the buoyancy effect; when the shear velocity decreases, the buoyancy effect is stronger than the shear effect.

Figure \ref{Fig-sm} shows the influence of different initial disturbance interface shapes on the early mechanism judgment of the coupled system. In the previous work, we have proved that, the two quantities, $L$ and $D_{3,1}$, always show a high correlation, especially in the early stage. Therefore, the simulation results of $D_{3,1}$ are only shown. In figures \ref{Fig-sm} (a) and (b), the black curve of the pure RTI system (g = 0.005) and the green curve of the pure KHI system with shear velocity $u_{0}=0.1$ intersect at a point, and remain approximately coincidence until this point, i.e., these two systems have the highest similar degree of medium mixing in the early stages. The critical shear velocity is $u_{C}=0.1$. In figures \ref{Fig-sm} (c) and (d), as the interface shape changes, the bilateral contact angle $\theta_{1}$ increases, the black curve of the pure RTI system deviates from the green curve, which indicates that the critical shear velocity $u_{C}$ increases. For the elliptic structure (Fig. \ref{Fig-sm} (d)), the critical shear velocity increases to $u_{C}=0.12$. Figures \ref{Fig-sm} (e) and (f) show similar trends. The black curve of the pure RTI system with $g = 0.005$ is approximately coincident with the red curve of the pure KHI system with shear velocity $u_{0}=0.08$ in the early stage, and then gradually deviates from it, and intersects with the green and blue lines (shear velocity $u_{0}=0.1, 0.12$). In general, for a given gravitational acceleration $g$, the critical shear velocity $u_{C}$ is related to the shape of the initial interface.

For the case where the KHI dominates at earlier time and the RTI dominates at later time, the evolution process can be roughly divided into two stages. In the previous work, we have pointed out that, before the transition point of the two stages, both $L$ and $D_{3,1}$ initially increase exponentially, and then increase linearly. Hence, the ending point of linear increasing $L$ or $D_{3,1}$ can work as a geometric or physical criterion for discriminating the two stages.

Figure \ref{Fig-sp} shows the effects of initial disturbance shapes on the transition points of the coupled system. The green box denotes the mean heat flux strength $D_{3,1}$, the red circle represents the $dD_{3,1}/dt$, and the blue vertical line marks the position of the transition point. As can be seen from the figure, for the initial disturbance interface whose bilateral contact angle $\theta_{1}$ is greater than $90^{o}$ (Figs. \ref{Fig-sp} (b), (c), and (d)), the delay of transition point is relatively slow and the magnitude is almost negligible, as the bilateral contact angle increases; For the inverted parabolic and elliptic interfaces (Figs. \ref{Fig-sp} (e) and (f)), the transition point has a relatively rapid and substantial delay, as the bilateral contact angle decreases; The inverted parabolic and elliptic interfaces (Figs. \ref{Fig-sp} (e) and (f)) have a longer linear growth stage, and the transition point is later than that of the parabolic and elliptic interfaces (Figs. \ref{Fig-sp} (c) and (d)).

%%%%%%%%%%%%%%%%%%%%%%%%%%%%%%%%%%%%%%%%
\begin{figure*}
\begin{center}
\includegraphics[width=0.75\textwidth]{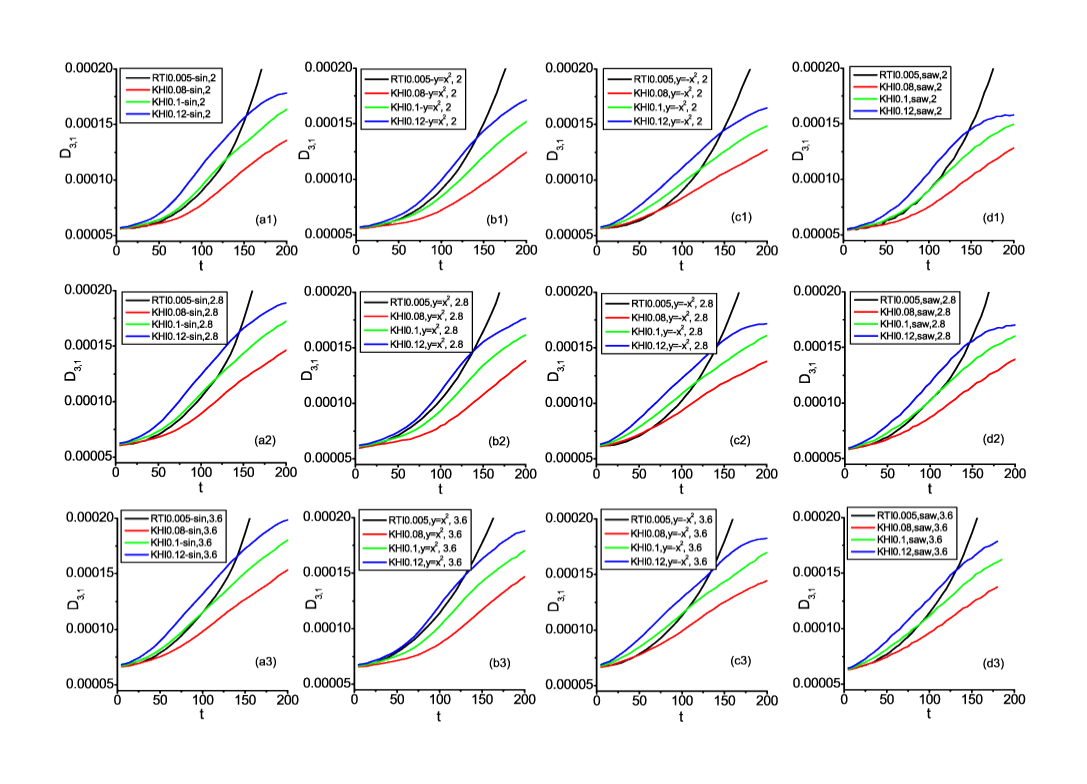}
\end{center}
\caption{(Color online) Effects of disturbance amplitude on the early main mechanism of the coupled system. (a), (b), (c), and (d) correspond to sinusoidal, parabolic, inverted parabolic and sawtooth disturbances, respectively. Rows 1 to 3 correspond to the initial amplitudes of $2$, $2.8$, $3.6$, respectively.}\label{Fig-am}
\end{figure*}
%%%%%%%%%%%%%%%%%%%%%%%%%%%%%%%%%%%%%%%%
%%%%%%%%%%%%%%%%%%%%%%%%%%%%%%%%%%%%%%%%
\begin{figure}[tbp]
\center\includegraphics*%
%[bbllx=14pt,bblly=15pt,bburx=319pt,bbury=310pt,width=0.48\textwidth]{Fig11.pdf}
[width=0.48\textwidth]{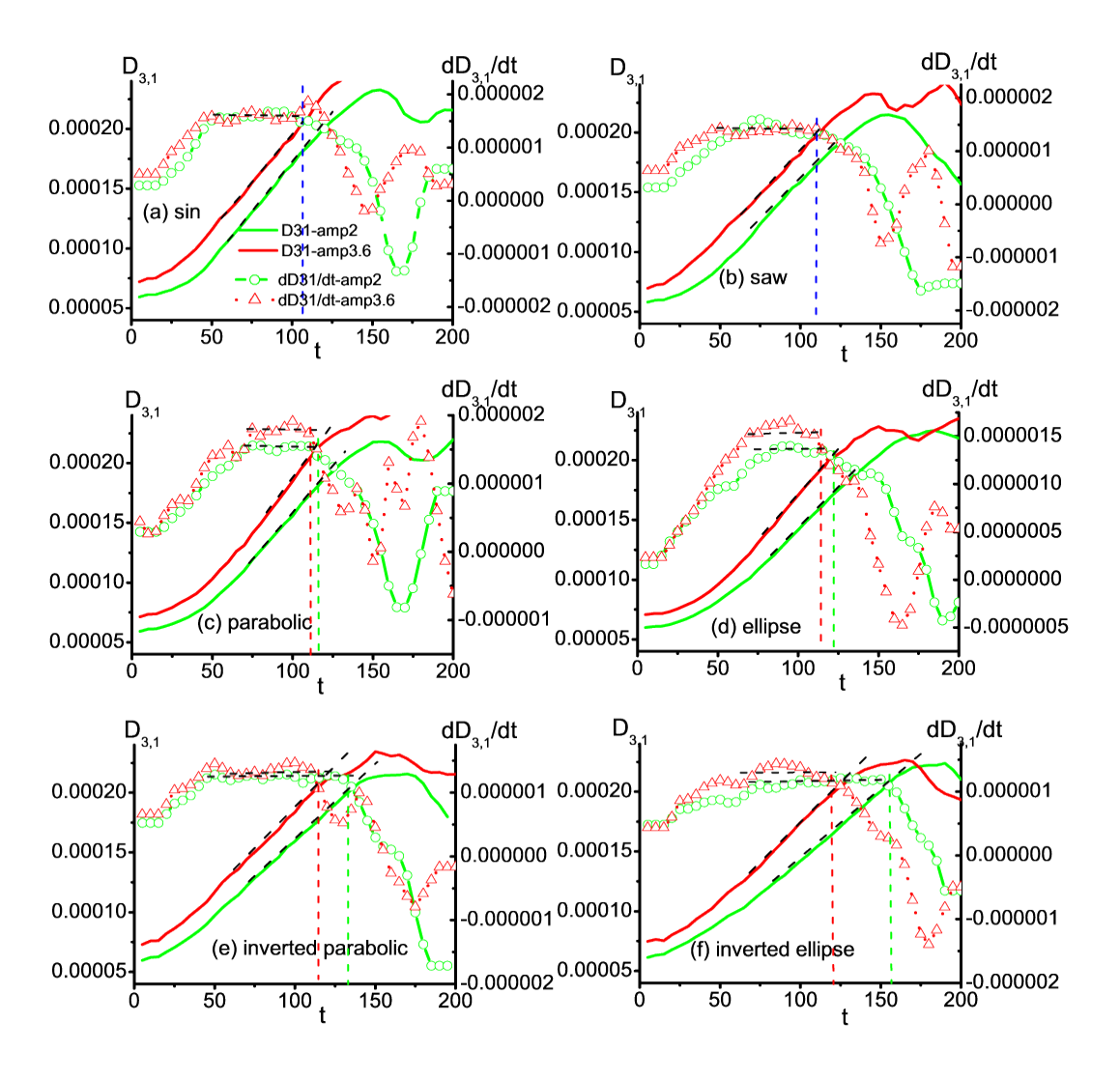}
\caption{(Color online) Effects of the initial disturbance amplitude on the transition point of the coupled system.}\label{Fig-ap}
\end{figure}
%%%%%%%%%%%%%%%%%%%%%%%%%%%%%%%%%%%%%%%%%

Figure \ref{Fig-am} shows the effect of disturbance amplitude on the early main mechanism of the coupled system. From top to bottom, the corresponding initial disturbance amplitudes are  $a_{0}=2$, $2.8$, and $3.6$, respectively. Figures \ref{Fig-am} (a), (b), (c), and (d) correspond to sinusoidal, parabolic, inverted parabolic and sawtooth disturbances, respectively. From the curves we can see that:

(i) For the sinusoidal disturbance, the contact angles ($\theta_{1}$ and $\theta_{2}$) of the interface are basically unchanged with the increase of amplitude, so the change of amplitude has no significant impact on the judgment of the early mechanism of the RTKHI system (Figs. \ref{Fig-am} (a1)-(a3)); For ellipse and inverted ellipse perturbation interfaces, there are consistent conclusions.

(ii) For the parabolic disturbance, with the increase of the disturbance amplitude, the middle contact angle $\theta_{2}$ almost remains unchanged, but the bilateral contact angle $\theta_{1}$ further increases, the black curve of pure RTI system with $g = 0.005$ gradually approaches the blue curve of pure KHI system with $u_{0} = 0.12$, that is, the critical velocity tends to increase with the increase of bilateral contact angle $\theta_{1}$ (Figs. \ref{Fig-am}(b1)-(b3)). This is consistent with the conclusion of Figs. \ref{Fig-sm}(c) and \ref{Fig-sm}(d).

(iii) For the inverted parabolic disturbance ($\theta_{2}=90^{o}$ and $\theta_{1}<90^{o}$), the amplitude increases and the bilateral contact angle $\theta_{1}$ decreases, but it does not affect the judgment of the early mechanism of the RTKHI system (Figs. \ref{Fig-am}(c1)-(c3)), and the images are similar to the system with inverted ellipse interface. This is also consistent with the conclusion of Figs. \ref{Fig-sm} (e) and (f).

(iv) For the sawtooth disturbance, the amplitude increases, the bilateral contact angle $\theta_{1}$ increases, but the middle contact angle $\theta_{2}$ decreases, and the critical velocity is basically unchanged due to the combined action of the two factors (Figs. \ref{Fig-am} (d1)-(d3)).

Figure \ref{Fig-ap} shows the effect of the initial disturbance amplitude on the transition point of the coupled system. The green and red curves correspond to the cases with $a_{0}=2$ and $a_{0}=3.6$, respectively. The virtual vertical lines indicate the positions of transition points. It can be seen from Fig. \ref{Fig-ap} that,
(i) For the RTKHI systems with different initial interfaces, the effects of amplitude on the transition point are not the same;
(ii) For the sinusoidal and sawtooth (Figs. \ref{Fig-ap} (a) and (b)), the increase of amplitude has no significant effect on the transition point (marked by blue vertical lines).
(iii) For the parabolic and ellipse interfaces (Figs. \ref{Fig-ap} (c) and (d)), as the amplitude increases, the transition points will be brought forward by a smaller amplitude. The results of $a_{0}=2$ and $a_{0}=3.6$ are marked with green and red vertical lines, respectively.
(iv) For the inverted parabolic and inverted ellipse perturbation interfaces Figs. \ref{Fig-ap} (e) and (f)), the increase of amplitude has a significant effect on the evolution process of the system and makes the transition point of the coupled system advance significantly. The effect of amplitude on the inverted ellipse interface is the greatest.

\section{Conclusions}

In this paper, we design six typical initial disturbance interfaces, and compare their effects on the RTI, KHI, and the coupled RTKHI systems. It is found that the initial perturbation has a great influence on the evolution of RTI. The sharper the interface, the faster the growth of bubbles or spikes. The influence of initial interface shape on KHI evolution can be ignored. Based on the mean heat flux strength $D_{3,1}$, the effects of initial interfaces on the coupled RTKHI system are investigated, and the researches focus on two aspects: (i) how the perturbed interface affects the transition point of the coupled system; (ii) how the perturbed interface affects the main mechanism in the early stage of the RTKHI system.

It is found that, in the coupled RTKHI system, for a given gravitational acceleration $g$, the magnitude of the critical shear velocity $u_{C}$ is closely related to the shape of the initial disturbance interface. The interface shape can be described by two contact angles, the bilateral contact angle $\theta_{1}$ and the middle contact angle $\theta_{2}$.
For the interface that $\theta_{1}$ and $\theta_{2}$ are basically constant, the critical velocity is basically constant. The increase of $\theta_{1}$ and the decrease of $\theta_{2}$ have opposite effects on the critical velocity. For the parabolic perturbation interface ($\theta_{2}=90^{o}$, $\theta_{1}>90^{o}$), the critical shear velocity increases as the increase of bilateral contact angle $\theta_{1}$. The ellipse perturbation interface is its limiting case. For the inverted parabolic and the inverted ellipse disturbance ($\theta_{2}=90^{o}$, $\theta_{1}<90^{o}$), the critical shear velocity is basically the same, which is less than that of the sinusoidal and sawtooth disturbances. The influence of inverted parabolic and inverted ellipse perturbations on the transition point of the RTKHI system is greater than that of other interfaces. When the amplitude is constant, the bilateral contact angle $\theta_{1}$ of the interface decreases, the transition point will be greatly delayed, and the transition point of the inverted ellipse structure system appears at the latest. For the same interface morphology, the disturbance amplitude increases, resulting in a shorter duration of the linear growth stage, and the transition point is greatly advanced.

\section*{Acknowledgments}

This work was supported by the Natural Science Foundation of Shandong Province (under Grant Nos. ZR2020MA061, ZR2019PA021), and Shandong Province Higher Educational Youth Innovation Science and Technology Program (under Grant No. 2019KJJ009), the National Natural Science Foundation of China (under Grant Nos. 11772064, 11875001, 12102397), CAEP Foundation (under Grant No. CX2019033), the opening project of State Key Laboratory of Explosion Science and Technology (Beijing Institute of Technology) (under Grant No. KFJJ21-16M), the China Postdoctoral Science Foundation (under Grant No. 2019M662521), Science Foundation of Hebei Province (under Grant No. A2021409001), ``Three, Three and Three Talent Project" of Hebei Province (under Grant No. A202105005).

\section*{Appendix}

$\hat{A}_{l}$ is the $l$th element of $\hat{\mathbf{A}}=(0,\cdots, 0, \hat{A}_{8}, \hat{A}_{9}, 0, \cdots, 0)$ and is a modification to the collision operator $\hat{S}_{lk}(\hat{f}_{k}-\hat{f}_{k}^{eq})$, where
\begin{align*}
& \hat{A}_{8}= (s_{T}/s_{v}-1)\rho Tu_{x} \left(2\frac{\partial u_{x}}{\partial x}-\frac{2}{b}\frac{%
\partial u_{x}}{\partial x}-\frac{2}{b}\frac{\partial u_{y}}{\partial y}%
\right)\notag \\
& \qquad +(s_{T}/s_{v}-1)\rho Tu_{y}\left(\frac{\partial u_{y}}{\partial x}+\frac{%
\partial u_{x}}{\partial y}\right)\text{,}
\end{align*}%
\begin{align*}
&  \hat{A}_{9}=(s_{T}/s_{v}-1)\rho Tu_{x} \left(\frac{\partial u_{y}}{\partial x}+\frac{\partial u_{x}}{%
\partial y}\right)\notag \\
& \qquad +(s_{T}/s_{v}-1)\rho Tu_{y}\left(2\frac{\partial u_{y}}{\partial y}-%
\frac{2}{b}\frac{\partial u_{x}}{\partial x}-\frac{2}{b}\frac{\partial u_{y}%
}{\partial y}\right)\text{,}
\end{align*}%
$s_{v}=s_{5}=s_{6}=s_{7}$, and $s_{T}=s_{8}=s_{9}$. This is to maintain the isotropy constraint of viscous
stress tensor and heat conductivity.

The transformation matrix and the corresponding equilibrium distribution functions in kinetic moment space are constructed according to the moment relations. Specifically,
the transformation matrix is
\begin{equation*}
\mathbf{M}=(m_{1},m_{2},\cdots ,m_{16})^{T}\text{,}\;
\end{equation*}
\begin{equation*}
m_{1}=1\text{,}\; m_{2}=v_{ix}\text{,}\; m_{3}=v_{iy}\text{,}\;
m_{4}=(v_{i\alpha }^{2}+\eta_{i}^{2})/2\text{,}\;
\end{equation*}
\begin{equation*}
m_{5}=v_{ix}^{2}\text{,}\; m_{6}=v_{ix}v_{iy}\text{,}\;
m_{7}=v_{iy}^{2}\text{,}\;
\end{equation*}
\begin{equation*}
m_{8}=(v_{i\beta}^{2}+\eta
_{i}^{2})v_{ix}/2\text{,}\; m_{9}=(v_{i\beta
}^{2}+\eta_{i}^{2})v_{iy}/2\text{,}\;
\end{equation*}
\begin{equation*}
m_{10}=v_{ix}^{3}\text{,}\; m_{11}=v_{ix}^{2}v_{iy}\text{,}\;
m_{12}=v_{ix}v_{iy}^{2}\text{,}\; m_{13}=v_{iy}^{3}\text{,}\;
\end{equation*}
\begin{equation*}
m_{14}=(v_{i\chi}^{2}+\eta _{i}^{2})v_{ix}^{2}/2\text{,}\;
m_{15}=(v_{i\chi }^{2}+\eta_{i}^{2})v_{ix}v_{iy}/2\text{,}\;
\end{equation*}
\begin{equation*}
m_{16}=(v_{i\chi }^{2}+\eta_{i}^{2})v_{iy}^{2}/2\text{.}\;
\end{equation*}

The corresponding equilibrium distribution functions in KMS are
\begin{equation*}
\hat{f}_{1}^{eq}=\rho \text{,}\; \hat{f}_{2}^{eq}=\rho
u_{x}\text{,}\; \hat{f}_{3}^{eq}=\rho u_{y}\text{,}\;
\hat{f}_{4}^{eq}=e\text{,}\;
\end{equation*}
\begin{equation*}
\hat{f}_{5}^{eq}=P+\rho u_{x}^{2}\text{,}\; \hat{f}_{6}^{eq}=\rho
u_{x}u_{y}\text{,}\; \hat{f}_{7}^{eq}=P+\rho u_{y}^{2}\text{,}\;
\end{equation*}
\begin{equation*}
\hat{f}_{8}^{eq}=(e+P)u_{x} \text{,}\;
\hat{f}_{9}^{eq}=(e+P)u_{y}\text{,}\; \hat{f}_{10}^{eq}=\rho
u_{x}(3T+u_{x}^{2})\text{,}\;
\end{equation*}
\begin{equation*}
\hat{f}_{11}^{eq}=\rho u_{y}(T+u_{x}^{2})\text{,}\;
\hat{f}_{12}^{eq}=\rho u_{x}(T+u_{y}^{2})\text{,}\;
\hat{f}_{13}^{eq}=\rho u_{y}(3T+u_{y}^{2})\text{,}\;
\end{equation*}
\begin{equation*}
\hat{f}_{14}^{eq}=(e+P)T+(e+2P)u_{x}^{2}\text{,}\;
\hat{f}_{15}^{eq}=(e+2P)u_{x}u_{y}\text{,}
\end{equation*}
\begin{equation*}
\hat{f}_{16}^{eq}=(e+P)T+(e+2P)u_{y}^{2}\text{,}
\end{equation*}
where pressure $P=\rho RT$ and energy $ e=b\rho RT/2+\rho u_{\alpha }^{2}/2$. $R$\ is the specific gas constant and $b$ is a constant related to the specific-heat-ratio $\gamma$ by $\gamma=(b+2)/b$.

%%%%%%%%%%%%%%%%%%%%%%%%%%%%%%%%%%%%%%%%%%%%%%%%%%

\hspace*{\fill} \\

\section*{References}
%\nocite{*}


\begin{thebibliography}{99}
\bibitem{DBM} Without causing misunderstanding, DBM is used as an abbreviation of discrete Boltzmann model/Modeling/Method.
\end{thebibliography}
\end{document}